\documentclass{amsart}
\usepackage[]{hyperref} 
\usepackage{float}
\usepackage{etoolbox}
\patchcmd{\subsection}{-.5em}{.5em}{}{}
\newtheorem{theorem}{Theorem}[section]

\theoremstyle{definition}

\newtheorem{example}{Example}[section]
\theoremstyle{remark}
\newtheorem{remark}{Remark}[section]
\theoremstyle{prop}

\theoremstyle{coro}
\newtheorem{corollary}[theorem]{Corollary}
\numberwithin{equation}{section}

\allowdisplaybreaks[1]
\begin{document}
\title[Incomplete Symmetric Orthogonal Polynomials of Finite Type]{Incomplete Symmetric Orthogonal Polynomials\\ of Finite Type Generated by a \\Generalized Sturm-Liouville Theorem}

\author[M. Masjed-Jamei]{Mohammad Masjed-Jamei$ ^{\dagger} $}

\author[Z. Moalemi]{Zahra Moalemi$ ^{\ddagger} $}

\author[N. Saad]{Nasser Saad$ ^{\dagger\ddagger} $}

\thanks{ $ ^{\dagger},^{\ddagger} $ Department of Mathematics, K.N.Toosi University of Technology, P.O.Box 16315-1618, Tehran, Iran.\\
$ ^{\dagger\ddagger} $ School of Mathematical and Computational Sciences, University of Prince Edward Island, Charlottetown, Canada. \\
{\em $ ^{\dagger} $ E-mail address}, M. Masjed-Jamei: mmjamei@kntu.ac.ir ; mmjamei@yahoo.com\\
URL: https://wp.kntu.ac.ir/mmjamei\\
{\em $ ^{\ddagger}$ E-mail address}, Z. Moalemi: zmoalemi@mail.kntu.ac.ir\\
{\em $ ^{\dagger\ddagger} $ E-mail address}, N. Saad: nsaad@upei.ca}


\date{}

\begin{abstract}
In a classical case, orthogonal polynomial sequences are in such a way that the $ n $th polynomial has the exact degree $n$. Such sequences are complete and form a basis of the space for any arbitrary polynomial. In this paper, we introduce some incomplete sets of finite orthogonal polynomials that do not contain all degrees but they are solutions of some symmetric generalized Sturm-Liouville problems. Although such polynomials do not possess all properties as in classical cases, they can be applied to functions approximation theory as we will compute their explicit norm square values. 
\end{abstract}


\maketitle

\section{Introduction}
\noindent We start with a second order differential equation in the form
\begin{equation}\label{EQ1.ODE}
\frac{d}{{dx}}\left( {k(x)\frac{{dy}}{{dx}}} \right) +\big(\lambda \rho(x)- q(x)\big)y=0,
\end{equation}
defined on an open interval, say $ (a,b) $, with the following boundary conditions
\begin{equation}\label{EQ2.boundary}
\begin{array}{l}
{\alpha _1}y(a)\, + {\beta _1}y'(a) = 0,\\[3mm]
{\alpha _2}y(b)\, + {\beta _2}y'(b) = 0,
\end{array}
\end{equation}
in which $ \alpha_{1},\alpha_{2} $ and $ \beta_{1},\beta_{2} $ are given constants and $ k(x)>0\,,\,k'(x)\,,\,q(x) $ and $ \rho(x)>0 $ in \eqref{EQ1.ODE} are assumed  to be continuous for all $ x\in [a,b] $.
\vskip0.1true in
The boundary value problem \eqref{EQ1.ODE}-\eqref{EQ2.boundary} is called a regular Sturm-Lioville problem and if one of the points $ a $ and $ b $ is singular (i.e. $ k(a)=0 $ or $ k(b)=0 $), it is called a singular Sturm-Liouville problem \cite{Arfken}. 
\vskip0.1true in

Let $ y_{n}(x) $ and $ y_{m}(x) $ be two solutions of equation \eqref{EQ1.ODE}. Following Sturm-Liouville theory \cite{Arfken, Niki}, these functions are orthogonal with respect to the positive weight function $ \rho(x) $ on $ (a,b) $ under the given conditions \eqref{EQ2.boundary}, i.e.
\begin{equation}\label{EQ3.defOrtho}
\int_a^b {\rho (x){y_n}(x){y_m}(x)\,dx}  = \left( {\int_a^b {\rho (x)y_n^2(x)\,dx} } \right){\delta _{n,m}},
\end{equation}
where
\[{\delta _{n,m}} = \left\{ \begin{array}{l}
0\,\,\,\,\,\,\,\,(n \ne m),\\
1\,\,\,\,\,\,\,\,(n = m).
\end{array} \right.\]

Many special functions in theoretical and mathematical physics are solutions of a regular or singular Sturm-Liouville problem satisfying the orthogonality condition \eqref{EQ3.defOrtho}. For instance, the associated Legendre functions, Bessel functions, trigonometric sequences related to Fourier analysis, ultraspherical functions and Hermite functions are particular solutions of a Sturm-Liouville problem \cite{Arfken, Chihara, Szego}. 
\vskip0.1true in
It is well known that the classical orthogonal polynomials of Jacobi, Laguerre and Hermite satisfy a second order differential equation of the form \cite{AlSalam, Andrews, Bochner}
\begin{equation}\label{EQ.hypDiff}
\sigma (x){y''_n}(x) + \tau (x){y'_n}(x) - {\lambda _n}{y_n}(x) = 0{\rm{ ,}}
\end{equation}
where 
\[ \sigma (x) = a{x^2} + bx + c \,\,\,\,\text{ and }\,\,\,\, \tau (x) = dx + e, \] 
are independent of $ n $ and 
\[ {\lambda _n} = n\big(d+(n-1)a\big),\]
is the eigenvalue depending on $ n=0,1,2,\ldots $. The Jacobi orthogonal polynomials for 
\[\sigma (x) = 1 - {x^2}\quad\text{and}\quad \tau (x) =  - (\alpha  + \beta  + 2)x + (\beta  - \alpha ),\] 
Laguerre polynomials for 
\[ \sigma (x) =x \quad\text{and}\quad \tau (x) =  \alpha  +1-x,\] 
and finally Hermit polynomials for 
\[ \sigma (x) =1 \quad\text{and}\quad \tau (x) = -2x ,\] 
are three types of polynomial solutions of equation \eqref{EQ.hypDiff} which are infinitely orthogonal \cite{Niki}. However, there are three other sequences of hypergeometric polynomials that are solutions of the above equation but finitely orthogonal with respect to the generalized T, inverse Gamma and F distributions \cite{Freund, Masjedjamei4}, respectively. 
\vskip0.1true in
Fortunately, most of the solutions of a regular or singular Sturm-Liouville problem are symmetric and have found interesting applications in physics and engineering \cite{Arfken, Szego}. If these functions are extended symmetrically while their orthogonality property is preserved, new applications in physics and engineering will naturally appear extending the previous ones. In this direction, the conditions under which the usual Sturm-Liouville problems with symmetric solutions are extended to a larger class have been recently studied in \cite{Masjedjamei35}. 
\begin{theorem}\label{th:1.1}
\cite{Masjedjamei35}. Let $ {\Phi _n}(x) = {( - 1)^n}{\Phi _n}(-x) $ be a sequence of independent symmetric functions that satisfies the differential equation 
\begin{equation}\label{EQ.ExtDiff}
A(x)\,{\Phi ''_n}(x) + B(x)\,{\Phi '_n}(x) + \left( {{\lambda _n}\,C(x) + D(x) + \left(\dfrac{1 - {{( - 1)}^n}}{2}\right)E(x)} \right)\,{\Phi _n}(x) = 0,
\end{equation}
where $ A(x)\,,\,\,B(x)\,,\,\,C(x)\,,\,\,D(x) $ and $ E(x) $ are real functions and $ \{ {\lambda _n}\} $ is a sequence of constants. If $ A(x)\,,\,\,(C(x)\, > 0),\,\,D(x) $ and $ E(x) $ are even functions and $ B(x) $ is odd then
\begin{equation*}
\int_{ - v}^v {\,{W^*}(x)\,{\Phi _n}(x)\,{\Phi _m}(x)\,dx}  = \left( {\int_{ - v}^v {\,{W^*}(x)\,\Phi _n^2(x)\,dx} } \right)\,{\delta _{n,m}},
\end{equation*}
where $ \,{W^*}(x) $ denotes the corresponding weight function as
\begin{equation}\label{EQ.ExtWeight}
\,{W^*}(x) = C(x)\exp \Big(\int {\frac{{B(x) - A'(x)}}{{A(x)}}\,dx} \Big) = \frac{{C(x)}}{{A(x)}}\exp \Big(\int {\frac{{B(x)}}{{A(x)}}\,dx} \Big).
\end{equation}
Of course, the weight function defined in \eqref{EQ.ExtWeight} must be positive and even on $ [ - v,v] $ and $ x=v $ must be a root of the function
\begin{equation*}
A(x)K(x) = A(x)\exp \Big(\int {\frac{{B(x) - A'(x)}}{{A(x)}}\,dx}\Big) = \exp \Big(\int {\frac{{B(x)}}{{A(x)}}\,dx}\Big),
\end{equation*}
i.e. $ A(v)\,K(v) = 0 $. Notice since $ K(x) = \dfrac{{W^*}(x)}{C(x)} $ is an even function, the relation $ A(-v)\,K(-v) = 0 $ follows automatically. 
\end{theorem}
According to the above theorem, many symmetric orthogonal functions can be generalized. For instance, a basic class of symmetric orthogonal polynomials with four free parameters is introduced in \cite{Masjedjamei36}. In this sense, a basic class of symmetric orthogonal functions with six parameters is introduced in \cite{Masjedjamei50} and their properties are studied there in detail, see also \cite{Masjedjamei37}. For other examples in this regard, we  refer to \cite{Masjedjamei39} and \cite{Masjedjamei42} as well as \cite{Masjedjamei114} where a new differential equation generalizing the differential
equation of Fourier trigonometric sequences possessing an orthogonal solution with respect to the constant weight function on $ [0,\pi] $ is introduced. It should be mentioned that further important cases still can be found that lead to a further motivation for more extensive investigations of orthogonal sequences. 
\vskip0.1true in
Recently, two sets of orthogonal polynomials known as incomplete symmetric orthogonal polynomials of Jacobi and Laguerre types have been introduced in \cite{Masjedjamei49} and \cite{Masjedjamei56}, respectively. Unlike the classical case, such sequences do not contain polynomials of every degree and therefore do not have all properties as in the classical cases. However, they can
be applied to functions approximation theory since their norm square values have been explicitly computed.
\vskip0.1true in

In this paper, by referring to the generalized Sturm-Liouville theorem \ref{th:1.1}, we introduce two incomplete symmetric orthogonal polynomials of finite type and obtain their orthogonality properties. For this purpose, in the next section we first  review three classes of finite orthogonal polynomials and study their orthogonality relations \cite{Masjedjamei4}.

\section{Three Finite Classes of Orthogonal Polynomials}

\subsection{First Finite Sequence}
The first finite sequence of classical orthogonal polynomials is defined for 
\[  \sigma (x) = {x^2} + x \quad\text{and} \quad \tau (x) = (2 - p)x + (1 + q),\] 
in \eqref{EQ.hypDiff} which leads to the differential equation
\begin{equation}\label{EQ.diF1}
({x^2} + x)\, {y''_n}(x) + \big((2 - p)x + q+1\big)\, {y'_n}(x) - n(n + 1 - p)\, {y_n}(x) = 0,
\end{equation}
having the explicit polynomial solution 
\begin{equation}\label{EQ.defM}
y_{n}(x)=M_n^{(p,\,q)}(x) = {( - 1)^n}n!\sum\limits_{k = 0}^{ n} {\,
\binom{p - n - 1}{k}
\left( \begin{array}{l}
q + n\\
n - k
\end{array} \right){{( - x)}^k}},
\end{equation}
where 
\[\binom{a}{k} = \frac{1}{{k!}}\prod\limits_{i = 0}^{k - 1} {(a - i)}.\] 
The polynomials \eqref{EQ.defM} are finitely orthogonal with respect to the weight function 
\[ {W_1}(x;p,q) = {x^q}{(1 + x)^{ - (p + q)}} ,\] 
on $ [0,\infty) $ if and only if 
\[p > 2\{ \max \,n\} \, + 1 \quad\text{and} \quad q>-1.\] 
Indeed, if the self-adjoint form of equation \eqref{EQ.diF1} is considered as
\[\Big({x^{1 + q}}{(1 + x)^{1 - p - q}}{y'_n}(x)\Big)' = n(n + 1 - p){x^q}{(1 + x)^{ - (p + q)}}{y_n}(x),\]
and for the index $ m $ as
\begin{equation}\label{EQ.SLM1}
\Big({x^{1 + q}}{(1 + x)^{1 - p - q}}{y'_m}(x)\Big)' = m(m + 1 - p){x^q}{(1 + x)^{ - (p + q)}}{y_m}(x),
\end{equation}
then multiplying by $ y_{m}(x) $ and $ y_{n}(x) $ in \eqref{EQ.SLM1} respectively and subtracting them gives
\begin{multline}\label{EQ.SLM2}
\left[ {\frac{{{x^{q + 1}}}}{{{{(1 + x)}^{p + q - 1}}}}\Big({y'_n}(x){y_m}(x) - {y'_m}(x){y_n}(x)\Big)} \right]_0^\infty  \\= ({\lambda _n} - {\lambda _m})\int_0^\infty  {\frac{{{x^q}}}{{{{(1 + x)}^{p + q}}}}M_n^{(p,q)}(x)M_m^{(p,q)}(x)\,dx} ,
\end{multline}
where $ {\lambda _n} = n(n + 1 - p) $.  Now, since 
\[\text{max deg}\,\,\{ {y'_n}(x){y_m}(x) - {y'_m}(x){y_n}(x)\}  = m + n - 1,\]
if 
\[ q>-1 \quad\text{and} \quad p>2N+1 \quad\text{for} \quad N=\max\lbrace m,n\rbrace,\] 
the left hand side of \eqref{EQ.SLM2} tends to zero and one gets
\begin{equation*}
\int_0^\infty  {\,\,\frac{{{x^q}}}{{{{(1 + x)}^{p + q}}}}M_n^{(p,q)}(x)M_m^{(p,q)}(x)\,dx}  = \,0\,\,\, \Leftrightarrow \,\,\,
m \ne n,\,\,p > 2N + 1\,\,\,\text{and}\,\,\,q >  - 1\,.
\end{equation*}
Hence, the complete orthogonality relation of the polynomials $ M_n^{(p,q)}(x) $ is given by
\begin{equation}\label{EQ.NormM}
\int_0^\infty  {\,\,\frac{{{x^q}}}{{{{(1 + x)}^{p + q}}}}M_n^{(p,q)}(x)M_m^{(p,q)}(x)\,dx}  = \,\frac{{n\,!\,(p - n - 1)!\,(q + n)!}}{{(p - 2n - 1)\,(p + q - n - 1)!}}\,{\delta _{n,m}},
\end{equation}
if and only if $ m,n = 0,1,2,...,N < \frac{{p - 1}}{2}$, $q >  - 1 $ and $ z!=\Gamma(z+1) $.\\

\bigskip

\subsection{Second Finite Sequence}

The second finite sequence is directly related to Bessel polynomials defined by 
\[ \sigma (x) = {x^2} \quad\text{and} \quad \tau (x)=2x+2 ,\] 
which were first studied by Krall and Frink in 1949 \cite{Krall}. Of course, they established a complex orthogonality of these polynomials on the unit circle. In 1973, the generalized Bessel polynomials were studied by Grosswald \cite{Grosswald}, which are indeed special solutions of equation \eqref{EQ.hypDiff} for 
\[ \sigma (x) = {x^2} \quad\text{and} \quad \tau (x) = (2 + \alpha )x + 2 ;\,\,\, \alpha \neq -2,-3, ...,\]  
leading to
\begin{equation*}
\bar B_n^{(\alpha )}(x) = {2^n}\sum\limits_{k = 0}^n {\left( \begin{array}{l}
n\\
k
\end{array} \right)\frac{{\Gamma (n + k + \alpha  + 1)}}{{\Gamma (2n + \alpha  + 1)}}{{\big(\frac{x}{2}\big)}^k}},
\end{equation*}
as the monic type of generalized Bessel polynomials.

\vskip0.1true in
The second finite sequence is defined in \cite{Masjedjamei4} for 
\[ \sigma (x) = {x^2} \quad\text{and}\quad \tau (x) = (2 + p)x + 1,\] 
leading to the differential equation
\begin{equation}\label{EQ.diF2}
{x^2}{y''_n}(x) + \big((2 + p)x + 1\big){y'_n}(x) - n(n + 1 + p){y_n}(x) = 0,
\end{equation}
with the polynomial solution
\begin{equation*}
y_{n}(x)=N_n^{(p)}(x) = {( - 1)^n}\sum\limits_{k = 0}^n {k!\binom{-p-n-1}{k}\binom{n}{n-k}{{( - x)}^k}} .
\end{equation*}
If equation \eqref{EQ.diF2} is written in the self-adjoint form 
\[\Big({x^{ p + 2}}\exp ( - \frac{1}{x}){y'_n}(x)\Big)' = n(n + 1+ p){x^{ p}}\exp ( - \frac{1}{x}){y_n}(x),\]
and for the index $ m $ as
\begin{equation}\label{EQ.SLN1}
\Big({x^{ p + 2}}\exp ( - \frac{1}{x}){y'_m}(x)\Big)' = m(m + 1 + p){x^{ p}}\exp ( - \frac{1}{x}){y_m}(x), 
\end{equation}
then multiplying by $ y_{m}(x)$ and $y_{n}(x) $ in \eqref{EQ.SLN1} respectively and subtracting them, one gets
\begin{multline}\label{EQ.SLN2}
\left[ {{x^{ p + 2}}{e^{ - \frac{1}{x}}}\Big({y'_n}(x){y_m}(x) - {y'_m}(x){y_n}(x)\Big)} \right]_0^\infty  \\
= ({\lambda _n} - {\lambda _m})\int_0^\infty  {{x^{ p}}{e^{ - \frac{1}{x}}}N_n^{(p)}(x)N_m^{(p)}(x)\,dx},
\end{multline}
where $ {\lambda _n} = n(n + 1 + p) $. Once again, since 
\[\max \deg  \{ {y'_n}(x){y_m}(x) - {y'_m}(x){y_n}(x)\}  = m + n - 1 ,\] 
the condition 
\[ p < -2N- 1\quad\text{for}\quad N = \max \{ m,n\},\] 
causes the left hand side of \eqref{EQ.SLN2} to tend to zero and therefore
\begin{equation*}
\int_0^\infty  {\,\,{x^{p}}{e^{ - \frac{1}{x}}}N_n^{(p)}(x)N_m^{(p)}(x)\,dx}  = \,0\,\,\, \Leftrightarrow \,\,\,
m \ne n,\,\,\text{and}\,\,p < -2N - 1.
\end{equation*}
Hence, the complete orthogonality relation is given by
\begin{equation}\label{EQ.NormN}
\int_0^\infty  {\,{x^{ p}}{e^{ - \frac{1}{x}}}N_n^{(p)}(x)N_m^{(p)}(x)\,dx}  =  {\frac{{n!\big(-p - (n + 1)\big)!}}{{-(p + 2n + 1)}}} \,{\delta _{n,m}},
\end{equation}
if and only if $ m,n = 0,1,2,...,N < -\frac{{p + 1}}{2} $.

\bigskip

\subsection{Third Finite Sequence}\label{secTFS3}

The third finite sequence  related to hypergeometric polynomials, which is finitely orthogonal with respect to the generalized T-student weight function 
\[ W_{3}(x;p,q) = {\left( 1+x^2 \right)^{ - p}}\exp (q\arctan x),\]
on $ (-\infty,\infty) $.

A particular case of equation \eqref{EQ.hypDiff} for 
\[ \sigma (x) = 1 + {x^2} \quad\text{and}\quad \tau (x) = (3 - 2p)x ,\] 
which is corresponding to the usual T student weight function, yields the equation
\begin{equation}\label{EQ.diF3}
(1 + {x^2}){y''_n}(x) + (3 - 2p)x\,{y'_n}(x) - n(n + 2 - 2p){y_n}(x) = 0,
\end{equation}
with the symmetric polynomial solution
\begin{equation*}
I_n^{(p)}(x) = n!\sum\limits_{k = 0}^{[\frac{n}{2}]} {{{( - 1)}^k}\left( \begin{array}{l}
p - 1\\
n - k
\end{array} \right)\left( \begin{array}{l}
\,n - k\\
\,\,\,\,k
\end{array} \right){{(2x)}^{n - 2k}}}, 
\end{equation*}
which are finitely orthogonal with respect to the usual T-student weight function
\[ W_{3}(x;p-\frac{1}{2},0)  = {(1 + {x^2})^{ - (p - \frac{1}{2})}}, \]
on $ (-\infty,\infty) $.

This claim can be easily proved by transforming \eqref{EQ.diF3} to a Sturm-Liouville equation and applying the same technique used for the first and second kind of finite classical orthogonal polynomials. This proccess eventually leads to
\begin{multline*}
\int_{ - \infty }^\infty  {(1+x^2)^{-(p-\frac{1}{2})}\,I_n^{(p)}(x)I_m^{(p)}(x)\,dx}  \\
= \frac{{n!\,{2^{2n - 1}}\sqrt \pi  \,{\Gamma ^2}(p)\,\Gamma (2p - 2n)}}{{(p - n - 1)\,\Gamma (p - n)\,\Gamma (p - n + 1/2)\,\Gamma (2p - n - 1)}}\,\,{\delta _{n,m}},
\end{multline*}
if and only if $ m,n = 0,1,2,...,N < p - 1 $.
\vskip0.1true in
Still a more extensive polynomial sequence which is finitely orthogonal on $ (-\infty,\infty) $ and generalizes the T-student distribution as its weight function. This nonsymmetric polynomial sequence is obtained by replacing 
\[ \sigma (x)=1+x^2 \quad\text{and}\quad  \tau (x)=2(1-p)x+q,\] 
in \eqref{EQ.hypDiff} leading to the equation 
\begin{equation}\label{EQ.diF32}
(1+{x^2} )\,y''_{n}(x) + \big( 2(1-p)x+q \big)\,{y'_n}(x)
 - n(n + 1 - 2p)\,{y_n}(x) = 0,
\end{equation}
with the polynomial solution
\begin{equation*}
J_n^{(p,q)}(x) = {( - i)^n}{(n + 1 - 2p)_n}\, 
 \sum\limits_{k = 0}^{ n} {\,\left( \begin{array}{l}
n\\
k
\end{array} \right)\,{}_2{F_1}\left( {\left.  \begin{array}{*{20}{c}}
{\begin{array}{*{20}{c}}
{ k- n,}&{p - n - iq/2}
\end{array}}\\
{2p - 2n}
\end{array}\right| {\,2} } \right)} \,{(-ix)^k},
\end{equation*}
in which 
\[ _2{F_1}\left( {\left.  \begin{array}{*{20}{c}}
{\begin{array}{*{20}{c}}
{a,}&{b}
\end{array}}\\
{c}
\end{array}\right| {x} } \right)=\sum\limits_{k=0}^{\infty}\dfrac{(a)_{k}(b)_{k}}{(c)_{k}}\dfrac{x^{k}}{k!} ,\] 
is the Gauss hypergeometric function \cite{Slater} and $ {(a)_k} = {\Gamma (a + k)}/{\Gamma (a)} $.
\vskip0.1true in
If equation \eqref{EQ.diF32} is written as
\begin{equation*}
{\Big({\left( {{1+x^2} } \right)^{1 - p}}\exp (q\arctan x){y'_n}(x)\Big)^\prime }
 = n(n + 1 - 2p){\left( 1+x^2 \right)^{ - p}}\exp (q\arctan x){y_n}(x),
\end{equation*}
then applying Sturm-Liouville theorem on $ (-\infty,\infty) $ gives
\begin{multline}\label{EQ.SLJ1}
\left[{\left( {{1+x^2} } \right)^{1 - p}}\exp (q\arctan x)({y'_n}(x){y_m}(x) - {y'_m}(x)\,{y_n}(x))\right]_{ - \infty }^\infty  \\
= \left( {{\lambda _n} - {\lambda _m}} \right)\int_{ - \infty }^\infty  {{{\left( 1+x^2 \right)}^{ - p}}\exp (q\arctan x)\,J_n^{(p,q)}(x)J_m^{(p,q)}(x)\,dx}, \,
\end{multline}
where $ {\lambda _n} = n(n + 1 - 2p) $. Since 
\[ \max \deg \lbrace {{y'_n}(x){y_m}(x) - {y'_m}(x){y_n}(x)} \rbrace = n + m - 1, \]  
if the conditions 
\[ p > N + \frac{1}{2}\,\,\,\text{for}\,\,\,N = \max \{ m,n\} \,\,\,{\rm{and}}\,\,q \in {\mathbb{R}},\] 
hold, the left hand side of \eqref{EQ.SLJ1} tends to zero and 
\begin{multline*}
\int_{ - \infty }^\infty  {(1+x^2)^{ - p}}\exp (q\arctan x)\,J_n^{(p,q)}(x)J_m^{(p,q)}(x)\,dx  = 0\\
 \Leftrightarrow 
m \ne n\,,\,p > N + \frac{1}{2}\,\,\text{and}\,\,
q \in {\mathbb{R}}.
\end{multline*}
Hence, the complete orthogonality relation is given by
\begin{multline}\label{EQ.NormJ}
\int_{ - \infty }^\infty  {{(1+x^2)^{ - p}}\exp (q\arctan x)\,J_n^{(p,q)}(x)\,J_m^{(p,q)}(x)\,dx} 
\\ = \Big(\frac{{n!\,\Gamma (2p - n)}}{{\Gamma (2p - 2n)}}\int_{ - \frac{\pi}{2}}^{\frac{\pi}{2}} {{{(\cos \theta   )}^{2p - 2n - 2}}\,{e^{q\theta }}d\theta }\Big )\,{\delta _{n,m}},
\end{multline}
if and only if $ m,n = 0,1,2,...,N < p - \frac{1}{2}\,\,\,\text{and}\,\,q \in {\mathbb{R}} $.
\vskip0.1true in
Note that \eqref{EQ.NormJ} will be simplified if $ 2p $ is a natural number, as the following Cauchy's formula \cite{Cauchy} holds
\[ \int_{-\frac{\pi}{2}}^{\frac{\pi}{2}}e^{st}\cos ^{r}t\,dt=\dfrac{\pi\,2^{-r}\Gamma(r+1)}{\Gamma(1+\frac{r+is}{2})\,\Gamma(1+\frac{r-is}{2})}. \]

\section{Incomplete Symmetric Orthogonal Polynomials of Finite Types}

\subsection{First Finite Sequence of Incomplete Type}

In this part, we introduce a finite sequence of incomplete symmetric orthogonal polynomials as a specific solution of a generalized Sturm-Liouville equation. In other words, by referring to the  theorem \ref{th:1.1} we can construct a differential equation of type \eqref{EQ.ExtDiff} whose solutions are orthogonal with respect to an even weight function, say $ |x|^{2a}\big(1+x^{2m}\big)^{b} $ on the symmetric interval $ (-\infty, \infty) $.

For this purpose, we first substitute
\begin{equation*}
g(x) = {x^\lambda }M_{n }^{(p ,q )}({x^\theta })\,\,\,\,\text{for}\,\,\,\lambda \,,\,\theta  \in {\mathbb{R}},
\end{equation*} 
into equation \eqref{EQ.diF1} to reach the differential equation
\begin{multline}\label{EQ.exdifM1}
{x^2}(1 +{x^\theta })\,g''(x) + x\Big( {-2\lambda +1+q\theta +\big(-2\lambda+1-(p-1)\theta\big)\,{x^\theta } } \Big)g'(x) \\
 + \Big( \lambda^{2}-\lambda q\theta+\Big(\lambda^{2}+\lambda\theta(p-1)-n\theta^{2}(n+1-p)\Big)x^{\theta} \Big)g(x)  = 0\,.
\end{multline}
For convenience, let 
\begin{equation*}
p = \frac{\theta-\beta-2\lambda+1}{\theta }\quad \,\,\,\,{\rm{and}}\quad\,\,\,\,q  =   \frac{\alpha+2\lambda-1}{\theta }.
\end{equation*} 
In this case, equation \eqref{EQ.exdifM1} changes to 
\begin{multline}\label{EQ.exdifM2}
{x^2}(1 + {x^\theta })\,g''(x) + x\left( {\beta\,{x^\theta } + \alpha} \right)g'(x) \\
+ \Big(\lambda (1-\lambda  -\alpha) -(\lambda+n\theta)(\lambda+n\theta+\beta-1)\,{x^\theta }  \Big)g (x)= 0.
\end{multline}
Now, by noting theorem \ref{th:1.1}, we define the following odd and even polynomial sequences
\begin{equation}\label{EQ.OEM}
\begin{array}{l}
{\Phi _{2n}}(x) = {x^{2s}}M_{n}^{\big(\frac{{2m-\beta-4s+1}}{{2m}},  \frac{\alpha+4s-1}{2m}\big)}({x^{2m}})  
\,\,\,\,\text{for}\,\,\, \lambda  = 2s\,,\,s \in \mathbb{N}\cup\{0\}\,{\rm{and}}\,\,\theta  = 2m\,,\,m \in {\mathbb{N}},\\[3mm]
\text{and}\\[3mm]
{\Phi _{2n + 1}}(x) = {x^{2r + 1}}M_{n}^{\big(\frac{{2m-\beta-4r-1}}{{2m}},  \frac{\alpha+4r+1}{2m}\big)}({x^{2m}})\,\,\,\,\text{for}\,\, \lambda  = 2r + 1\,,\,r \in { \mathbb{N}\cup\{0\}}\,{\rm{and}}\,\theta  = 2m\,,\,m \in {\mathbb{N}}.
\end{array}
\end{equation} 
If we take
\begin{equation}\label{EQ.sigman}
\sigma_{n}=\dfrac{1-(-1)^{n}}{2},
\end{equation}
the polynomial sequence $ {\Phi _n}(x) $ defined in \eqref{EQ.OEM} can be written in a unique form as
\begin{equation*}
{\Phi _n}(x) =\big({x^{2s}} + ({x^{2r + 1}} - {x^{2s}})\,{\sigma _n}\big)\,M_{[\frac{n}{2}]}^{\big(\frac{2m-\beta-2s-2r+(-1)^{n}(2r-2s+1)}{2m}, \frac{\alpha+2s+2r+(-1)^{n}(2s-2r-1)}{2m}\big)}({x^{2m}})\,.
\end{equation*}
According to definitions \eqref{EQ.OEM} and differential equation \eqref{EQ.exdifM2}, $ {\Phi _{2n}}(x) $ should satisfy the equation
\begin{multline*}
{x^2}(1 + {x^{2m}})\,{\Phi''_{2n}}(x) + x\,\left( {\beta{x^{2m}} + \alpha} \right){\Phi'_{2n}}(x)\\
 + \Big(2s(1-\alpha-2s ) -2(s+mn)(2s+2mn+\beta-1)\,{x^{2m}}  \Big){\Phi _{2n}}(x) = 0\,,
\end{multline*}
and $ {\Phi _{2n+1}}(x) $ should satisfy
\begin{multline*}
{x^2}(1 + {x^{2m}})\,{\Phi''_{2n + 1}}(x) + x\left( {\beta{x^{2m}} + \alpha} \right){\Phi'_{2n + 1}}(x)\\
+ \Big( - (2r + 1)(2r + \alpha) -(2r+2mn+1)(2r+2mn+\beta)\,{x^{2m}} \Big){\Phi _{2n + 1}}(x) = 0\,.
\end{multline*}
Combining these two equations finally gives 
\begin{multline}\label{EQ.OEcombineM}
{x^2}(1 + {x^{2m}})\,{\Phi''_n}(x) + x\left( {\beta x^{2m} + \alpha} \right){\Phi'_n}(x) + \\
\Big(2s(1-\alpha-2s)-\big((2r+1)(2r+\alpha)+2s(1-\alpha-2s)\big)\sigma_{n}\\
-\big((2s+mn)(2s+mn+\beta-1)+(2r-2s-m+1)(2r+2s-m+\beta+2mn)\sigma_{n}\big) x^{2m}\Big){\Phi _n}(x) = 0\,,
\end{multline}
which is a special case of the generalized Sturm-Liouville equation \eqref{EQ.ExtDiff}. 
\vskip0.1true in

Also, the weight function corresponding to equation \eqref{EQ.OEcombineM} takes the form 
\begin{equation}\label{EQ.ExtWeightM}
W(x) = \dfrac{x^{2m}}{x^{2}(1+x^{2m})}\exp \Big(\int {\dfrac{\alpha+\beta x^{2m}}{x(1+x^{2m})}\,dx} \Big) 
= K\,{x^{2m + \alpha - 2}}{(1 + {x^{2m}})^{\frac{\beta-\alpha}{2m}-1}}.
\end{equation}
Without loss of generality, we can assume that $ K=1 $ and since $ W(x) $ must be positive, the weight function \eqref{EQ.ExtWeightM} can be considered as $ {\left| x \right|^{\,2a}}{(1 + {x^{2m}})^b}  $ for $ 2a = 2m + \alpha - 2 $ and $ b =  -1+({\beta-\alpha})/({2m}) $.

\begin{corollary}
Suppose in the generic equation \eqref{EQ.ExtDiff} that
\begin{equation*}
\begin{array}{l}
A(x) = {x^2}(1 + {x^{2m}}),\\
B(x) =  2x\,\big( (a+mb+1){x^{2m}} +a - m + 1\big),\\
C(x) = {x^{2m}} > 0,\\
D(x) = D = -2s\big(2(a+s-m)+1\big),\\
E(x) = E = 2s(2s + 2a - 2m + 1) - 2(2r + 1)(r + a - m + 1),
\end{array}
\end{equation*}
and
\[{\lambda _n} =-(2s+mn)(2s+mn+2mb+2a+1)-(2r-2s-m+1)(2r+2s+2mb+2a+2mn-m+2)\sigma_{n} ,\]
where $  {\sigma _n} = ({{1 - {{( - 1)}^n}}})/{2} $ and $ a,b,m,r,s $ are free parameters. Then the differential equation 
\begin{multline}\label{EQ.mainDifM}
{x^2}(1 + {x^{2m}})\,{\Phi ''_n}(x) +2x\,\big( (a+mb+1){x^{2m}} +a - m + 1\big)\,{\Phi '_n}(x) \\
+ \Big( {\lambda _n}{x^{2m}}+E\sigma_n +D \Big)\,{\Phi _n}(x) = 0,
\end{multline}
has an incomplete polynomial solution in the form 
\begin{align}
&\Phi _n^{(r,s)}(x;a,b,m) = \big({x^{2s}} + ({x^{2r + 1}} - {x^{2s}})\,{\sigma _n}\big)\,M_{[\frac{n}{2}] }^{(u,v)}({x^{2m}})\label{EQ.DefPhi}\\
&=\big({x^{2s}} + ({x^{2r + 1}} - {x^{2s}})\,{\sigma _n}\big)  {( - 1)}^{[\frac{n}{2}]}[\frac{n}{2}]!\sum\limits_{k = 0}^{[\frac{n}{2}]} {\,
\binom{u - [\frac{n}{2}] - 1}{k}
\binom{v + [\frac{n}{2}]}{[\frac{n}{2}] - k}
{{( -1)}^{k} \,x^{2mk}}},\nonumber
\end{align}
in which 
\[u=\dfrac{1}{m}\Big({m(1-b)-(a+s+r+1)+(-1)^{n}(r-s+\frac{1}{2})}\Big),\]
and
\[v=\dfrac{1}{m}\Big({a+s+r-m+1+(-1)^{n}(s-r-\frac{1}{2})}\Big).\]
This solution satisfies the finite orthogonality relation
\begin{multline}\label{EQ.ExtOrthoM}
\int_{ - \infty}^{\infty} {{x^{2a}}{{(1 + {x^{2m}})}^b}\Phi _n^{(r,s)}(x;a,b,m)\Phi _k^{(r,s)}(x;a,b,m)\,dx}  \\
= \left( {\int_{ - \infty}^{\infty} {{x^{2a}}{{(1 + {x^{2m}})}^b}{{\left( {\Phi _n^{(r,s)}(x;a,b,m)} \right)}^2}dx} } \right)\,{\delta _{n,k}}.
\end{multline}  

\end{corollary}

To compute the norm square value of \eqref{EQ.ExtOrthoM} we directly use the orthogonality relation \eqref{EQ.NormM} so that for $ n=2j $ we have
\begin{align}
{\mathcal{N}_{2j}} &= \int_{ - \infty}^{\infty} {{x^{2a}}{{(1+{x^{2m}})}^b}{{\left( {\Phi _{2j}^{(r,s)}(x;a,b,m)} \right)}^2}dx} \nonumber \\
&= \int_{ - \infty}^{\infty} {{x^{2a + 4s}}{{(1 + {x^{2m}})}^b}{{\left( {M_{j}^{(1-b-\frac{2a+4s+1}{2m},-1+\frac{2a+4s+1}{2m})}({x^{2m}})} \right)}^2}dx}\nonumber  \\
 &= \frac{1}{m}\int_0^{\infty} {{t^{\frac{{2a + 4s + 1 - 2m}}{{2m}}}}{{(1+ t)}^b}{{\left( {M_{j }^{(1-b-\frac{2a+4s+1}{2m},-1+\frac{2a+4s+1}{2m})}(t)} \right)}^2}dt} \nonumber \\
  &= \dfrac{j!\,\big(-(b+j+\frac{2a+4s+1}{2m})\big)!\,\big(j-1+\frac{2a+4s+1}{2m}\big)!}{-\big(m(b+2j)+a+2s+\frac{1}{2}\big)\,\big(-(b+j+1)\big)!},\label{EQ.ExtNormM1}
\end{align}
and for $ n=2j+1 $ we have
\begin{align}
{\mathcal{N}_{2j + 1}} &= \int_{ - \infty}^{\infty} {{x^{2a}}{{(1 + {x^{2m}})}^b}{{\left( {\Phi _{2j + 1}^{(r,s)}(x;a,b,m)} \right)^2}}dx} \nonumber \\
&= \int_{ -\infty}^{\infty} {{x^{2a + 4r + 2}}{{(1 + {x^{2m}})}^b}{{\left( {M_{j}^{(1-b-\frac{2a+4r+3}{2m},-1+\frac{2a+4r+3}{2m})}({x^{2m}})} \right)^2}}dx}  \nonumber \\
&=\frac{1}{m}\int_{0}^{\infty} {{t^{\frac{{2a + 4r + 3 - 2m}}{{2m}}}}{{(1 + t)}^b}{{\left( {M_{j}^{(1-b-\frac{2a+4r+3}{2m},-1+\frac{2a+4r+3}{2m})}({x^{2m}})(t)} \right)^2}}dt}\nonumber  \\
& = \dfrac{j!\,\big(-(b+j+\frac{2a+4r+3}{2m})\big)!\,\big(j-1+\frac{2a+4r+3}{2m}\big)!}{-\big(m(b+2j)+a+2r+\frac{3}{2}\big)\,\big(-(b+j+1)\big)!}.\label{EQ.ExtNormM2}
\end{align}
By combining both relations \eqref{EQ.ExtNormM1} and \eqref{EQ.ExtNormM2} we eventually obtain
\begin{multline}\label{MainNormM}
{\mathcal{N}_n} =\\
\dfrac{(\frac{n-\sigma_{n}}{2})!\,\Big(-\big(\frac{n}{2}+b+\frac{2a+4s+1}{2m}+\frac{4r-4s-m+2}{2m}\sigma_{n}\big)\Big)!\,\Big(\frac{n}{2}-1+\frac{2a+4s+1}{2m}+\frac{4r-4s-m+2}{2m}\sigma_{n}\Big)!}{-\Big(m(n+b)+a+2s+\frac{1}{2}+(2r-2s-m+1)\sigma_{n}\Big)\,\Big(-\big(\frac{n-\sigma_{n}}{2}+b+1\big)\Big)!}.
\end{multline}

To determine the parameters constraint in \eqref{MainNormM}, we write equation \eqref{EQ.mainDifM} in a self-adjoint form as
\begin{equation*}
\Big(x^{2a-2m+2}(1 + {x^{2m}})^{b+1}\,{\Phi '_n}(x)\Big)'  \\
+ \Big({\lambda _n}{x^{2m}}  + E\sigma_n +D \Big)\,x^{2a-2m}(1+x^{2m})^{b}{\Phi _n}(x) = 0,
\end{equation*}
to obtain
\begin{multline}\label{EQ.SelfM}
\left[ x^{2a-2m+2}(1 + {x^{2m}})^{b+1}\,\Big(\Phi '_n(x)\Phi _k(x)-\Phi '_k(x)\Phi _n(x)\Big)\right] _{-\infty}^{\infty}\\
=(\lambda_{n}-\lambda_{k})\int_{ - \infty}^{\infty} {{x^{2a}}{{(1 + {x^{2m}})}^b}\Phi _n(x)\Phi _k(x)\,dx}.
\end{multline}
Since $ \Phi _n(x) $ defined in \eqref{EQ.DefPhi} is a symmetric sequence of degree at most $ mn+2s+(2r-2s-m+1)\sigma_{n} $, we have
\[\text{max deg}\,\left\lbrace  {\Phi '_n}(x){\Phi _k}(x) - {\Phi '_k}(x){\Phi _n}(x)\right\rbrace   = mn+mk+4s-1+(2r-2s-m+1)(\sigma_{n}+\sigma_{k}).\]
Hence, for $ N = \max \{n,k\} $ the condition 
\begin{equation}\label{EQ.PM1}
2mN+2a+2mb+4s+1+(2r-2s-m+1)(\sigma_{n}+\sigma_{k})<0,
\end{equation}
causes the left hand side of \eqref{EQ.SelfM} to tend to zero. Four cases may occur for $ n $ and $ k $ in inequality \eqref{EQ.PM1} which are respectively
\begin{equation*}
i)\left\{ \begin{array}{l}
n = 2i\\
k = 2j + 1
\end{array} \right.\qquad ii)\left\{ \begin{array}{l}
n = 2i + 1\\
k = 2j
\end{array} \right.\qquad iii)\left\{ \begin{array}{l}
n = 2i\\
k = 2j
\end{array} \right.\qquad iv)\left\{ \begin{array}{l}
n = 2i + 1\\
k = 2j + 1.
\end{array} \right.\,
\end{equation*}
Replacing each above cases in \eqref{EQ.PM1} leads to
\begin{equation}\label{EQ.PM2}
\begin{array}{l}
\qquad\qquad N<-\dfrac{1}{m}\big(a+2s+\frac{1}{2}\big)-b,\\[3mm]
\qquad\qquad N<-\dfrac{1}{m}\big(a+2r+\frac{3}{2}\big)-b+1,\\[3mm]
\text{and}\\[3mm]
\qquad\qquad N<-\dfrac{1}{m}\big(a+s+r+1\big)-b+\dfrac{1}{2}.
\end{array}
\end{equation}

Moreover, due to positivity of any norm square value, the right hand side of \eqref{MainNormM} shows that the finite orthogonality \eqref{EQ.ExtOrthoM} is valid if the following conditions hold:
\begin{equation}\label{EQ.PM3}
\begin{array}{l}
m(n-\sigma_{n}+b)+a+2s+\frac{1}{2}+(2r-2s+1)\sigma_{n}<0,\\[3mm]
mn+2a+4s+1+(4r-4s-m+2)\sigma_{n}>0,\\[3mm]
n-\sigma_{n}<-2b,
\end{array}
\end{equation}
provided that $ m\in\mathbb{N} $. Again, by considering different even and odd values of $ n $ in above conditions and taking
\begin{multline*}
C_{r,s}^{(a,b,m)}=\\
\min\left\lbrace -\dfrac{1}{2m}\big(2a+4s+1\big)-b,\,-\dfrac{1}{2m}\big(2a+4r+3\big)-b+1,\,-\dfrac{1}{m}\big(a+s+r+1\big)-b+\dfrac{1}{2} \right\rbrace,
\end{multline*}
conditions \eqref{EQ.PM2} and \eqref{EQ.PM3} altogether yield the following corollary.
\begin{corollary}
The finite orthogonality relation \eqref{EQ.ExtOrthoM} is valid for $ n,k=0,1,\ldots,N<C_{r,s}^{(a,b,m)} $ if and only if 
\[|2a+4s+1|<-2mb,\,\,\,|2a+4r+3|<-2mb,\,\,\,b<0,\,\,\,m\in\mathbb{N},\,\,\,r,s\in\mathbb{N}\cup\{0\}\,\,\,\text{and}\,\,\,(-1)^{2a}=1.\]
\end{corollary}

\begin{example}
If we set $ m=1 $ in \eqref{EQ.DefPhi}, then the incomplete polynomials
\[ \left\lbrace \Phi _n^{(r,s)}(x;a,b,1) =\big({x^{2s}} + ({x^{2r + 1}} - {x^{2s}})\,{\sigma _n}\big)\,M_{[\frac{n}{2}] }^{(u,v)}({x^{2}})\right\rbrace _{n=0}^{N}, \]
for 
\[u=-(a+b+r+s)+(-1)^{n}(r-s+\dfrac{1}{2}), \]
and
\[ v=a+r+s+(-1)^{n}(s-r-\dfrac{1}{2}), \]
are derived such that for 
\[ N<C_{r,s}^{(a,b,1)}=\min\left\lbrace -(a+b+2s+\frac{1}{2}),\,-(a+b+2r)-\frac{1}{2},\,-(a+b+r+s)-\frac{1}{2} \right\rbrace  ,\]  they satisfy the orthogonality relation
\begin{multline*}
\int_{ - \infty}^{\infty} {{x^{2a}}{{(1 + {x^{2}})}^b}\Phi _n^{(r,s)}(x;a,b,1)\Phi _k^{(r,s)}(x;a,b,1)\,dx}  \\
=
\dfrac{(\frac{n-\sigma_{n}}{2})!\,\Big(-\big(\frac{n+1}{2}+a+b+2s+(2r-2s+\frac{1}{2})\sigma_{n}\big)\Big)!\,\Big(\frac{n-1}{2}+a+2s+(2r-2s+\frac{1}{2})\sigma_{n}\Big)!}{-\Big(n+a+b+2s+\frac{1}{2}+2(r-s)\sigma_{n}\Big)\,\Big(-\big(\frac{n-\sigma_{n}}{2}+b+1\big)\Big)!}
 \,{\delta _{n,k}},
\end{multline*}
if and only if 
\[|2a+4s+1|<-2b,\,\,\,|2a+4r+3|<-2b,\,\,\,b<0,\,\,\,r,s\in\mathbb{N}\cup\{0\}\,\,\,\text{and}\,\,\,(-1)^{2a}=1.\]
For instance, taking $ r=s=0 $, $ a=1 $ and $ b=-200 $ we have the finite set of polynomials
\[ \left\lbrace \Phi _n^{(0,0)}(x;1,-200,1) =\big(1 + (x - 1)\,{\sigma _n}\big)\,M_{[\frac{n}{2}] }^{(-201+\frac{(-1)^{n}}{2},1-\frac{(-1)^{n}}{2}}({x^{2}})\right\rbrace _{n=0}^{N}, \]
which is of course complete and satisfies the finite orthogonality relation
\begin{multline*}
\int_{ - \infty}^{\infty} {{x^{2}}{{(1 + {x^{2}})}^{-200}}\Phi _n^{(0,0)}(x;1,-200,1)\Phi _k^{(0,0)}(x;1,-200,1)\,dx}  \\
=
\dfrac{(\frac{n-\sigma_{n}}{2})!\,\Big(199-\frac{n+1}{2}-\frac{1}{2}\sigma_{n}\Big)!\,\Big(\frac{n+1}{2}+\frac{1}{2}\sigma_{n}\Big)!}{\Big(199-n-\frac{1}{2}\Big)\,\Big(199-\frac{n-\sigma_{n}}{2}\Big)!}
 \,{\delta _{n,k}},
\end{multline*}
for $ n,k=0,1,2,\ldots,198 $.
\end{example}

\bigskip

\subsection{Second Finite Sequence of Incomplete Type}

Once again, by referring to the  theorem \ref{th:1.1}, we can construct a differential equation of type \eqref{EQ.ExtDiff} whose solutions are orthogonal with respect to another even weight function, say $ |x|^{2a}\exp\big(-x^{-2m}\big) $ on the symmetric interval $ (-\infty, \infty) $.

To reach this goal, let us substitute
\begin{equation*}
h(x) = {x^\lambda }N_{n }^{(p)}({x^\theta })\,\,\,\,\text{for}\,\,\,\lambda \,,\,\theta  \in {\mathbb{R}},
\end{equation*} 
into equation \eqref{EQ.diF2} to reach the differential equation
\begin{multline}\label{EQ.exdifN1}
x^{\theta+2}h''(x) + x\Big( \big(-2\lambda+(p+1)\theta+1\big){x^\theta }+\theta  \Big)h'(x) \\
 + \left( -\lambda\theta+\Big(\lambda\big(\lambda-(p+1)\theta\big)-n\theta^{2}(n+p+1)\Big)x^{\theta} \right)h(x)  = 0\,.
\end{multline}
For convenience, let
\begin{equation*}
p=\dfrac{\alpha+2\lambda-1}{\theta}-1.
\end{equation*} 
In this case, equation \eqref{EQ.exdifN1} changes to 
\begin{equation}\label{EQ.exdifN2}
x^{\theta+2}h''(x) + x\big( \alpha{x^\theta }+\theta  \big)h'(x) 
 + \Big( -\lambda\theta-(\lambda+n\theta)(\lambda+n\theta+\alpha-1)\,x^{\theta} \Big)h(x)  = 0\,.
\end{equation}
By considering theorem \ref{th:1.1}, we can define the following odd and even polynomial sequences
\begin{equation}\label{EQ.OEN}
\begin{array}{l}
{\Psi_{2n}}(x) = {x^{2s}}N_{n}^{(\frac{\alpha+4s-1}{2m}-1)}({x^{2m}})
\,\,\,\,\text{for}\,\,\, \lambda  = 2s\,,\,s \in \mathbb{N}\cup\{0\}\,{\rm{and}}\,\,\theta  = 2m\,,\,m \in {\mathbb{N}},\\[3mm]
\text{and}\\[3mm]
{\Psi _{2n + 1}}(x) = {x^{2r + 1}}N_{n}^{(\frac{\alpha+4r+1}{2m}-1)}({x^{2m}})\,\,\,\,\text{for}\,\, \lambda  = 2r + 1\,,\,r \in \mathbb{N}\cup\{0\}\,{\rm{and}}\,\theta  = 2m\,,\,m \in {\mathbb{N}}.
\end{array}
\end{equation} 
Note that the polynomial sequence $ {\Psi _n}(x) $ defined in \eqref{EQ.OEN} can be written in a unique form as
\begin{equation*}
{\Psi _n}(x) =\big({x^{2s}} + ({x^{2r + 1}} - {x^{2s}})\,{\sigma _n}\big)\,N_{[\frac{n}{2}]}^{\big(\frac{\alpha+2s+2r+(-1)^{n}(2s-2r-1)}{2m}-1\big)}({x^{2m}}),
\end{equation*}
where $ \sigma_{n} $ is defined in \eqref{EQ.sigman}.

According to definitions \eqref{EQ.OEN} and differential equation \eqref{EQ.exdifN2}, $ {\Psi _{2n}}(x) $ should satisfy the equation
\begin{multline*}
x^{2m+2}{\Psi''_{2n}}(x) + x\big( \alpha{x^{2m} }+2m  \big){\Psi'_{2n}}(x)\\
 + \Big( -4ms-\,2\big(s+mn\big)\big(2s+2nm+\alpha-1\big){x^{2m}}  \Big){\Psi _{2n}}(x) = 0\,,
\end{multline*}
and $ {\Psi _{2n+1}}(x) $ should satisfy
\begin{multline*}
x^{2m+2}{\Psi''_{2n+1}}(x) + x\big( \alpha{x^{2m} }+2m  \big){\Psi'_{2n+1}}(x)\\
 + \Big( -2m(2r+1)-\big(2r+2mn+1\big)\big(2r+2mn+\alpha\big){x^{2m}}  \Big){\Psi _{2n+1}}(x) = 0\,.
\end{multline*}
Combining these two equations finally gives a special case of the generalized Sturm-Liouville equation \eqref{EQ.ExtDiff} as
\begin{multline}\label{EQ.OEcombineN}
x^{2m+2}{\Psi''_{n}}(x) + x\big( \alpha{x^{2m} }+2m  \big){\Psi'_{n}}(x)\\
 + \Big( -4ms-\big(2m(2r+1)-4ms\big)\sigma_{n} -\big((2s+mn)(2s+mn+\alpha-1)\\
 +(2r-2s-m+1)(2r+2s-m+\alpha+2mn)\sigma_{n}\big) {x^{2m}}\Big){\Psi _{n}}(x) = 0\,.
\end{multline}
Also, the weight function corresponding to \eqref{EQ.OEcombineN} takes the form 
\begin{equation}\label{EQ.ExtWeightN}
W(x) = \dfrac{x^{2m}}{x^{2m+2}}\exp \Big(\int {\dfrac{\alpha x^{2m+1}+2mx}{x^{2m+2}}\,dx} \Big) 
= K^{*}\,{x^{\alpha - 2}}\exp\big(-x^{-2m}\big).
\end{equation}
We can again assume that $ K^{*}=1 $ and since $ W(x) $ must be positive, the weight function \eqref{EQ.ExtWeightN} can be considered as $ {|x|^{2a}}\exp\big(-x^{-2m}\big)  $ for $ 2a = \alpha - 2 $.

\begin{corollary}
Suppose in the generic equation \eqref{EQ.ExtDiff} that
\begin{equation*}
\begin{array}{l}
A(x) = x^{2m+2},\\
B(x) =  2x\big((a+1)x^{2m}+m\big),\\
C(x) = {x^{2m}} > 0,\\
D(x) = D = -4ms,\\
E(x) = E = 2m(2s-2r-1),
\end{array}
\end{equation*}
and
\begin{equation*}
{\lambda _n} =-(2s+mn)(2s+mn+2a+1)
 -(2r-2s-m+1)(2r+2s-m+2a+2mn+2)\sigma_{n},
\end{equation*}
where $  {\sigma _n} = \frac{{1 - {{( - 1)}^n}}}{2} $ and $ a,m $ and $ r,s $ are free parameters. Then the differential equation 
\begin{equation}\label{EQ.mainDifN}
x^{2m+2}{\Psi''_{n}}(x) + 2x\big((a+1)x^{2m}+m\big){\Psi'_{n}}(x)
 + \big(\lambda_{n}x^{2m}+E\sigma_{n}+D\big){\Psi _{n}}(x) = 0,
\end{equation}
has an incomplete polynomial solution in the form 
\begin{align}
&\Psi _n^{(r,s)}(x;a,m) = \big({x^{2s}} + ({x^{2r + 1}} - {x^{2s}})\,{\sigma _n}\big)\,N_{[\frac{n}{2}] }^{\big(\frac{a+s+r+1+(-1)^{n}(s-r-\frac{1}{2})}{m}-1\big)}({x^{2m}})\label{EQ.DefPsi}\\[3mm]
&=\big({x^{2s}} + ({x^{2r + 1}} - {x^{2s}})\,{\sigma _n}\big) {( - 1)^{{[\frac{n}{2}]}}}\sum\limits_{k = 0}^{{[\frac{n}{2}]}} {k!\binom{-\frac{a+s+r+1+(-1)^{n}(s-r-\frac{1}{2})}{m}-{[\frac{n}{2}]}}{k}\binom{{[\frac{n}{2}]}}{{[\frac{n}{2}]}-k}{{( -1)}^k\,x^{2mk}}},\nonumber
\end{align}
which satisfies the finite orthogonality relation
\begin{multline}\label{EQ.ExtOrthoN}
\int_{ - \infty}^{\infty} {{x^{2a}}\exp({-x^{-2m}})\Psi _n^{(r,s)}(x;a,m)\Psi _k^{(r,s)}(x;a,m)\,dx}  \\
= \left( {\int_{ - \infty}^{\infty} {{x^{2a}}\exp({-x^{-2m}}){{\left( {\Psi _n^{(r,s)}(x;a,m)} \right)}^2}dx} } \right)\,{\delta _{n,k}}.
\end{multline}  

\end{corollary}

To compute the norm square value of \eqref{EQ.ExtOrthoN} we directly use the orthogonality relation \eqref{EQ.NormN} so that for $ n=2j $ we have
\begin{align}
{\mathcal{N}_{2j}} &= \int_{ - \infty}^{\infty} {{x^{2a}}e^{-\frac{1}{x^{2m}}}{{\left( {\Psi _{2j}^{(r,s)}(x;a,m)} \right)}^2}dx} \nonumber \\
&= \int_{ - \infty}^{\infty} {{x^{2a + 4s}}e^{-\frac{1}{x^{2m}}}{{\left( {N_{j}^{(\frac{2a+4s+1}{2m}-1)}({x^{2m}})} \right)}^2}dx}\nonumber  \\
 &= \frac{1}{m}\int_0^{\infty} {{t^{(\frac{{2a + 4s + 1 }}{{2m}}-1)}} e^{-\frac{1}{t}} {{\left( {N_{j }^{(\frac{2a+4s+1}{2m}-1)}(t)} \right)}^2}dt} = \dfrac{j!\,\big(-\frac{2a+4s+1}{2m}-j\big)!}{-\big(a+2s+\frac{1}{2}+2mj\big)},\label{EQ.ExtNormN1}
\end{align}
and for $ n=2j+1 $ we have
\begin{align}
{\mathcal{N}_{2j+1}} &= \int_{ - \infty}^{\infty} {{x^{2a}}e^{-\frac{1}{x^{2m}}}{{\left( {\Psi _{2j+1}^{(r,s)}(x;a,m)} \right)}^2}dx} \nonumber \\
&= \int_{ - \infty}^{\infty} {{x^{2a + 4r+2}}e^{-\frac{1}{x^{2m}}}{{\left( {N_{j}^{(\frac{2a+4r+3}{2m}-1)}({x^{2m}})} \right)}^2}dx}\nonumber  \\
 &= \frac{1}{m}\int_0^{\infty} {{t^{(\frac{{2a + 4r + 3 }}{{2m}}-1)}} e^{-\frac{1}{t}} {{\left( {N_{j }^{(\frac{2a+4r+3}{2m}-1)}(t)} \right)}^2}dt} = \dfrac{j!\,\big(-\frac{2a+4r+3}{2m}-j\big)!}{-\big(a+2r+\frac{3}{2}+2mj\big)},\label{EQ.ExtNormN2}
\end{align}
By combining both relations \eqref{EQ.ExtNormN1} and \eqref{EQ.ExtNormN2} we eventually obtain
\begin{equation}\label{MainNormN}
{\mathcal{N}_n} =\dfrac{(\frac{n-\sigma_{n}}{2})!\,\Big(-\frac{2a+4s+mn+1}{2m}+\frac{4s-4r+m-2}{2m}\,\sigma_{n}\Big)!}{-\Big(a+2s+mn+\frac{1}{2}+(2r-2s-m+1)\sigma_{n}\Big)}.
\end{equation}

To determine the parameters constraint in \eqref{MainNormN}, we write equation \eqref{EQ.mainDifN} in a self-adjoint form as
\begin{equation*}
\Big(x^{2a+2}\exp\big(-x^{-2m}\big)\,{\Psi '_n}(x)\Big)'  \\
+ \Big({\lambda _n}{x^{2m}}  + E\sigma_n +D \Big)\,x^{2a-2m}\exp\big(-x^{-2m}\big){\Psi _n}(x) = 0,
\end{equation*}
to obtain
\begin{multline}\label{EQ.SelfN}
\left[ x^{2a+2}\exp\big(-x^{-2m}\big)\,\Big(\Psi '_n(x)\Psi _k(x)-\Psi '_k(x)\Psi _n(x)\Big)\right] _{-\infty}^{\infty}\\
=(\lambda_{n}-\lambda_{k})\int_{ - \infty}^{\infty} {{x^{2a}}\exp\big(-x^{-2m}\big)\Psi _n(x)\Psi _k(x)\,dx}.
\end{multline}
Since $ \Psi _n(x) $ defined in \eqref{EQ.DefPsi} is a symmetric sequence of degree at most $ mn+2s+(2r-2s-m+1)\sigma_{n} $, we have
\[\text{max deg}\,\left\lbrace  {\Psi '_n}(x){\Psi _k}(x) - {\Psi '_k}(x){\Psi _n}(x)\right\rbrace   = mn+mk+4s-1+(2r-2s-m+1)(\sigma_{n}+\sigma_{k}).\]
Hence, for $ N = \max \{n,k\} $ the condition 
\begin{equation}\label{EQ.PN1}
2mN+2a+4s+1+(2r-2s-m+1)(\sigma_{n}+\sigma_{k})<0,
\end{equation}
causes the left hand side of \eqref{EQ.SelfN} to tend to zero. Again, four cases may occur for $ n $ and $ k $ in inequality \eqref{EQ.PN1} which eventually lead to
\begin{equation*}\label{EQ.PN2}
\begin{array}{l}
\qquad\qquad N<-\dfrac{1}{2m}\big(2a+4s+1\big),\\[3mm]
\qquad\qquad N<-\dfrac{1}{2m}\big(2a+4r+3\big)+1,
\end{array}
\end{equation*}
and
\[\qquad\qquad N<-\dfrac{1}{m}\big(a+s+r+1\big)+\dfrac{1}{2}.\]
Note that the above conditions also guarantee the positivity of the right hand side of \eqref{MainNormN} for $ m\in\mathbb{N} $.
Also, it is clear that we should have
\[ 2a+4s+1<0 ,\quad 2a+4r+3<2m\quad\text{and}\quad 2(a+s+r+1)<m.\]
Finally by taking
\begin{equation*}
C_{r,s}^{(a,m)}=
\min\left\lbrace -\dfrac{1}{2m}\big(2a+4s+1\big),\,-\dfrac{1}{2m}\big(2a+4r+3\big)+1,\,-\dfrac{1}{m}\big(a+s+r+1\big)+\dfrac{1}{2} \right\rbrace,
\end{equation*}
we reach the following corollary.
\begin{corollary}
The finite orthogonality relation \eqref{EQ.ExtOrthoN} is valid for $ n,k=0,1,\ldots,N<C_{r,s}^{(a,m)} $ if and only if 
\[2a+4s+1<0 ,\,\, 2a+4r+3<2m,\,\, 2(a+s+r+1)<m,\,\,\,m\in\mathbb{N},\,\,r,s\in\mathbb{N}\cup\{0\}\,\,\text{and}\,\,(-1)^{2a}=1.\]
\end{corollary}

\begin{example}
As a particular example, here we introduce incomplete symmetric polynomials orthogonal with respect to the weight function $ x^{-102}\exp(-x^{-4}) $ on $ (-\infty,\infty) $ which are represented as
\begin{equation*}
\Psi_{n}^{(r,s)}(x;-51,2)=\big({x^{2s}} + ({x^{2r + 1}} - {x^{2s}})\,{\sigma _n}\big)\,N_{[\frac{n}{2}] }^{\big(-26+\frac{s+r+(-1)^{n}(s-r-\frac{1}{2})}{2}\big)}({x^{4}}).
\end{equation*}
These polynomials satisfy the orthogonality relation
\begin{multline*}
\int_{ - \infty}^{\infty} {{x^{-102}}\exp({-x^{-4}})\Psi _n^{(r,s)}(x;-51,2)\Psi _k^{(r,s)}(x;-51,2)\,dx}  \\
=\dfrac{(\frac{n-\sigma_{n}}{2})!\,\Big(25-s-\frac{2n-1}{4}+(s-r)\,\sigma_{n}\Big)!}{\frac{99}{2}-\Big(2s+2n+(2r-2s-1)\sigma_{n}\Big)} \,{\delta _{n,k}},
\end{multline*} 
which is valid if 
\[ n,k<C_{r,s}^{(-51,2)}=\min\left\lbrace \frac{101}{4}-s,\frac{103}{4}-r,\frac{51-(s+r)}{2}\right\rbrace,\]
 for $ r,s\in \lbrace 0,1,\ldots,25\rbrace $.\\
It is clear that $ \Psi _n^{(r,s)}(x;-51,2) $ are incomplete symmetric polynomials with the degrees  $ 2n+2s+(2r-2s-1)\sigma_{n} $,
for $ n=0,1,\ldots, N<C_{r,s}^{(-51,2)}$.\\
For instance, if we take $ r=3 $ and $ s=1 $ then $  C_{3,1}^{(-51,2)}=\frac{91}{4} $ and the finite set of symmetric polynomials
\[ \left\lbrace \Psi_{n}^{(3,1)}(x;-51,2)=\big({x^{2}} + ({x^{7}} - {x^{2}})\,{\sigma _n}\big)\,N_{[\frac{n}{2}] }^{-(24+\frac{5}{2}(-1)^{n})}({x^{4}})\right\rbrace _{n=0}^{N\leq 22} ,\]
is orthogonal with respect to the weight function $ x^{-102}\exp(-x^{-4}) $ on $ (-\infty,\infty) $. Degrees of such incomplete polynomials are $ \lbrace 2,7,6,11,\ldots,47,46\rbrace $, respectively.
\end{example}

\bigskip

\begin{remark}
Note that it is not possible to derive a finite sequence of incomplete type corresponding to the third finite classical orthogonal polynomials $ J_n^{(p,q)}(x) $ or $ I_n^{(p)}(x) $ as their orthogonality interval, i.e. $ (-\infty,\infty) $, is already symmetric. Indeed, in this case the change of variable $ t=x^{\theta} $ where $ \theta=2m $, for $ x\in(-\infty,\infty) $  will not yield the symmetric interval $ (-\infty,\infty) $. Therefore, the conditions of the main theorem \ref{th:1.1} are not fulfilled and the norm square value can not be computed. 
\end{remark}

\bigskip

\section*{Acknowledgments}
\noindent The work of the first author has been supported by the {\em{Alexander von Humboldt Foundation}} under the grant number: Ref 3.4 - IRN - 1128637 - GF-E. 
Also a partial financial support of this work to the third author under Grant No. GP249507 from the Natural Sciences and Engineering Research
Council of Canada is gratefully acknowledged.

\end{document}